\documentclass[11pt,a4paper]{article}

\usepackage{graphicx}
\title{Plasma heating in the very early and decay phases of solar flares}

\begin{document}
\maketitle


\begin{center}
\author{R. Falewicz$^1$, M. Siarkowski$^2$ and P. Rudawy$^1$}
\end{center}
\author{ {\footnotesize $^1$Astronomical Institute, University of Wroc{\l}aw, 51-622 Wroc{\l}aw, \\ul. Kopernika 11, Poland}}
\begin{center}
\author{ \emph{\footnotesize falewicz@astro.uni.wroc.pl; rudawy@astro.uni.wroc.pl}}
\end{center}
\author{ {\footnotesize$^2$Space Research Centre, Polish Academy of Sciences, 51-622 Wroc{\l}aw, \\ul. Kopernika 11, Poland}}
\begin{center}
\author{\emph{\footnotesize ms@cbk.pan.wroc.pl} }
\end{center}
\author{\footnotesize The Astrophysical Journal (accepted, March 2011)}





\begin{abstract}

In this paper we analyze the energy budgets of two single-loop solar flares under the assumption that non-thermal electrons are the only source of plasma heating during all phases of both events. The flares were observed by the Ramaty High Energy Solar Spectroscopic Imager (\emph{RHESSI}) and  Geostationary Operational Environmental Satellite (\emph{GOES}) on September 20,  2002 and March 17, 2002, respectively. For both investigated flares we derived the energy fluxes contained in non-thermal electron beams from the  \emph{RHESSI} observational data constrained by observed \emph{GOES} light-curves.
We showed that energy delivered by non-thermal electrons was fully sufficient to fulfil the energy budgets of the plasma during the pre-heating and impulsive phases of both flares as well as during the decay phase of one of them. We concluded that in the case of the investigated flares there was no need to use any additional ad-hoc heating mechanisms other than heating by non-thermal electrons.

\end{abstract}


{Sun: flares --- Sun: X-rays, gamma rays }



\section{Introduction}

The common flare model, well consistent with their main observational signatures, comprises an energy transfer from a magnetic energy release region toward the chromosphere by non-thermal electron beams, an induced evaporation of the chromospheric plasma and vigorous radiation in a whole range of the electromagnetic spectrum. In particular, both hard X-ray (HXR) and soft X-ray (SXR) emissions are related to a flux of the non-thermal electrons (NTEs), while the HXR emission is directly excited in a bremsstrahlung process by the NTEs, and the SXR emission, thermal in origin, is related to the energy deposited by NTEs in the plasma.

Such a model, despite its overall elegancy and self-consistency, does lead to important considerations concerning the importance of various auxiliary processes of the energy transport, a total energy budget and time relations between SXR and HXR emissions (e.g.: Dennis 1988; Dennis \& Zarro 1993; McTierman et al.  1999; Falewicz et al. 2009). It happens quite often that the SXR emission in flares starts a few minutes earlier than the HXR emission, the maximum of the SXR emission occurs much later after the end of the HXR event and decay-times of the SXR emission are much longer than the decay-times estimated using radiative and conductive losses of energy.

The cadence of the SXR and HXR emissions during an initial phase of the flares has been investigated by several authors (Machado et al., 1986, Dennis, 1988, Schmahl et al., 1989, Veronig et al., 2002a) as well as pre-heating processes (Heyvearts et al., 1977, Li et al., 1987, Waren, 2006), while Battaglia et al., (2009) investigated an alternative mechanism of conducted-driven evaporation.

Similar problems and questions concern the relations between SXR and HXR fluxes after flare maxima. Based on a sample of 1114 flares Veronig et al. (2002b) have found that 270 events ($\sim 25\%$ of the analyzed sample) and the SXR maximum occurred distinctly after the end of the HXR emission.  The fact that the SXR emission is still increasing, although the HXR emission, i.e. the electron input, had apparently already stopped, provides the strong impression that an additional agent (besides the HXR emitting electrons) is contributing to the energy input and prolonging the heating and/or evaporation. Possible physical processes invoked by various authors as additional heating agents are, for instance, thermal conduction (Zarro \& Lemen 1988; Yokoyama \& Shibata 1998; Czaykowska et al. 2001), accelerated protons (Simnett 1986; Plunkett \& Simnett 1994), plasma waves (Petrosian 1994; Lee et al. 1995) and DC-electric fields (McDonald et al. 1999). Most of the previous works on the hydro-dynamic simulation of solar flares have not been able to account for the evolution of the observed emission (e.g., Peres et al.  1987; Mariska \& Zarro 1991). In fact hydro-dynamic simulations indicated that high-density flare plasma cools rapidly, while observed soft X-ray emission from solar flares usually persists for many hours. It suggests that some heating is present well into the decay phase (e.g., Serio et al. 1991).

\begin{figure}[t]
\includegraphics[angle=0,scale=0.40]{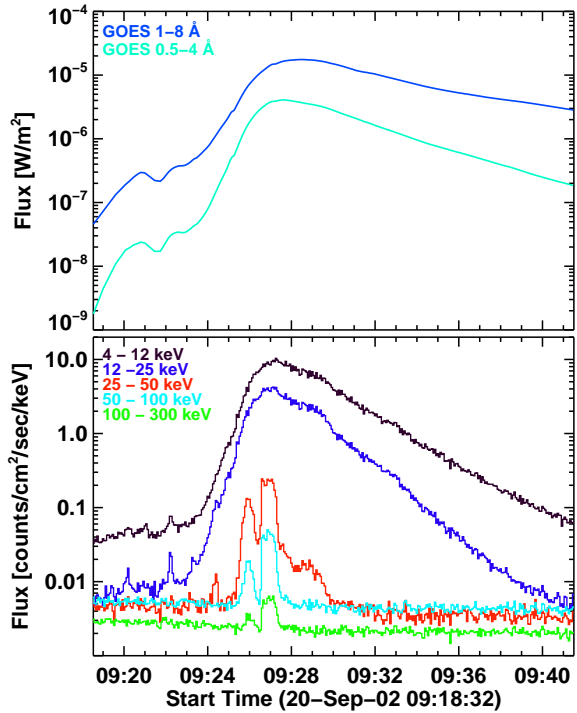}
\vspace{0.02 cm}
\caption{\emph{GOES} X-ray 0.5-4 \AA\   and 1-8 \AA\   light curves (upper panel) and \emph{RHESSI} light curves of five energy bands between 4 and 300 keV (lower panel) taken during the M1.8 \emph{GOES} class solar flare on September 20, 2002.
}

\end{figure}

In our previous paper (Siarkowski et al., 2009; thereafter called Paper I) we have shown, due to an unprecedented high sensitivity of the \emph{RHESSI} detectors (Lin et al. 2002) and using a numerical model of a single-loop flare, that for the M1.8 \emph{GOES} class solar flare on September 20, 2002 an early SXR emission observed prior to the impulsive phase could be fully explained by electron beam-driven evaporation and without any additional ad hoc assumptions.  In this modeled event all energy necessary to explain the observed SXR emission could be derived from observed HXR spectra, i.e. was delivered by NTEs. In the present paper we essentially extended our investigations of this flare assuming that electron beam-driven evaporation is the main heating mechanism acting not only during pre- but also during post-impulsive phases of the solar flares. We present here an extended modeling of the whole September 20, 2002 flare from pre-impulsive up to late gradual phases. We also present calculations made for pre-impulsive and impulsive phases of the M4.0 \emph{GOES} class event observed in AR NOAA 9871 on March 17, 2002. Unfortunately, the event showed at least a double-loop structure during the decay phase and thus our model was not relevant for its proper modeling at that time. It is worth stressing that both flares were observed by \emph{RHESSI} without the activation of the attenuators, thus they were very convenient for 1D hydro-dynamic numerical modeling while there were not any discontinuities in parameters describing spectra.

The paper is structured in the following way. Section 2 describes the analyzed events. Section 3 presents the details of the HXR spectra fitting, numerical modeling of the flares, and the results. The discussion and conclusions are presented in Section 4.

\section{Observations}

\begin{figure*}[t]
\includegraphics[angle=0,scale=0.45]{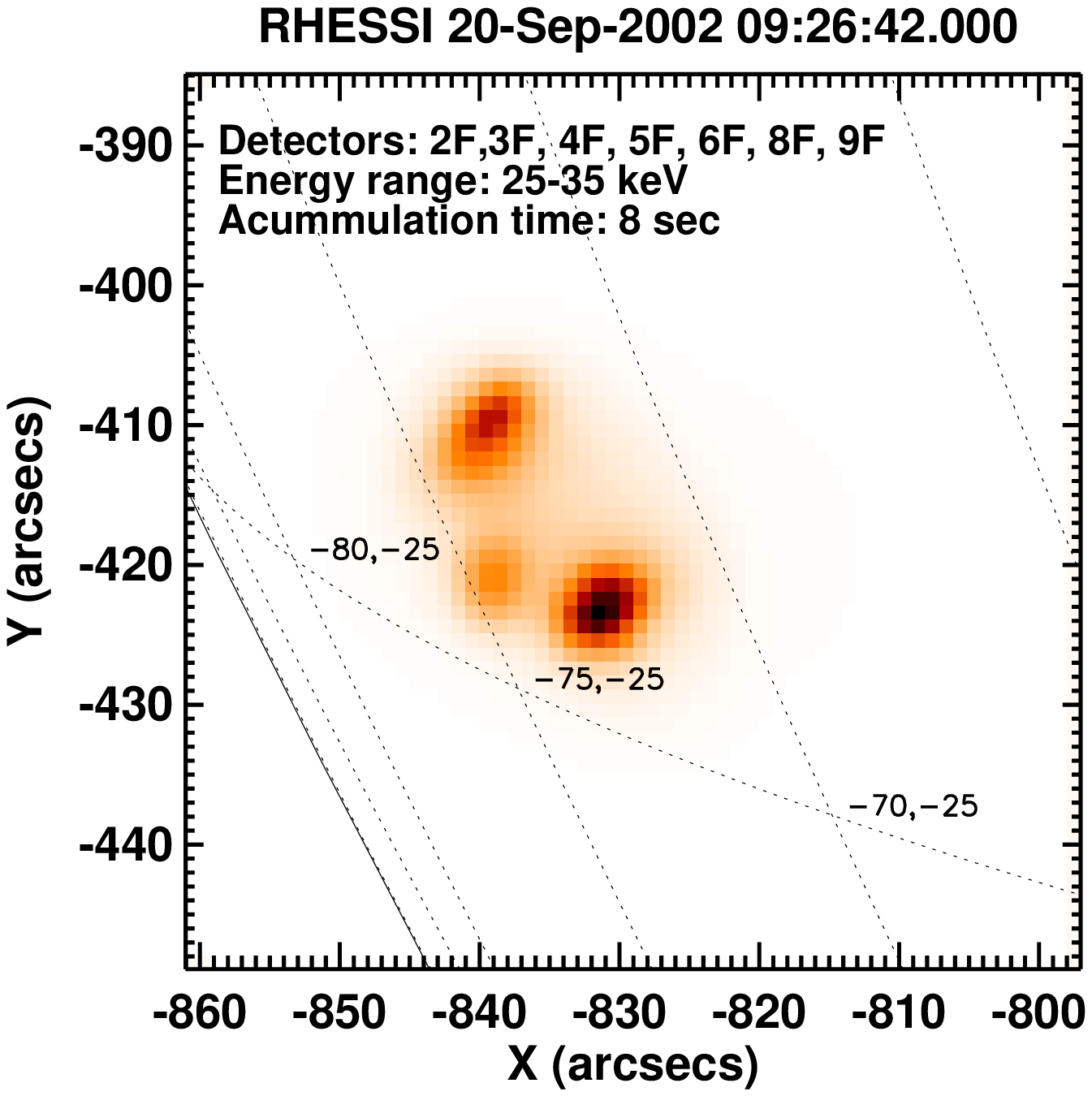}
\includegraphics[angle=0,scale=0.45]{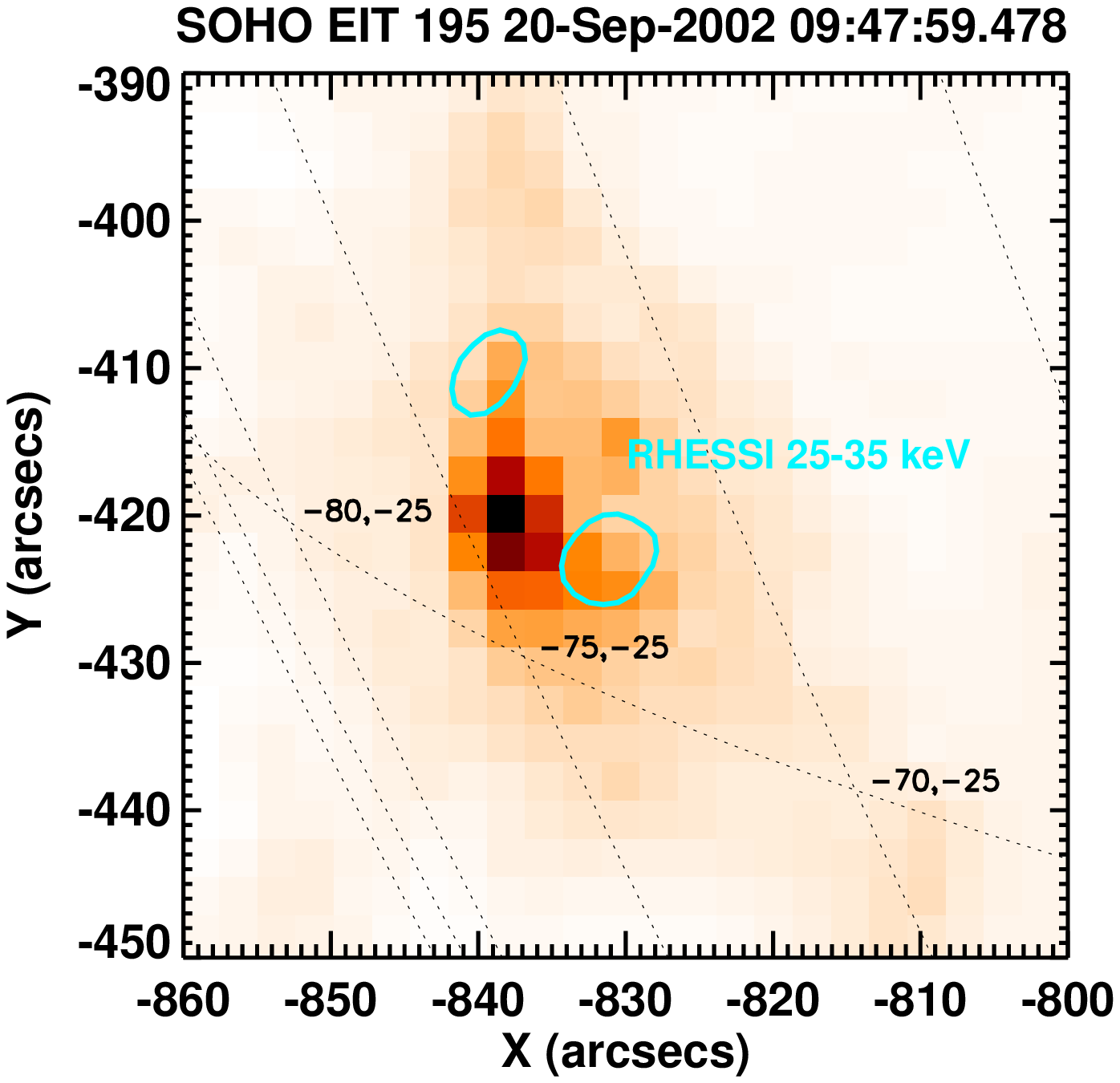}
\vspace{0.02 cm}
\caption{Images of the M1.8 \emph{GOES} class solar flare on September 20, 2002. Left panel: an image restored using the PIXON method in 25-35 keV energy band, signal was accumulated between 09:26:42 UT and 09:26:50 UT (see also Paper I).  Right panel: \emph{SOHO/EIT} 195 \AA\ image taken at 09:47:59 UT (gray scale) over-plotted with \emph{RHESSI} 25-35 keV image registered at 09:26:42 UT (contours).
}

\end{figure*}


For our work we selected two solar flares with a simple single-loop initial structure, convenient for numerical modeling. The flares were observed on March 17, 2002 and September 20, 2002. Both events were recorded by the Ramaty High Energy Solar Spectroscopic Imager (\emph{RHESSI}) satellite without the activation of the attenuators, thus the investigated spectra do not have any discontinuities (Lin et al.  2002; Hurford et al.  2002; Smith et al.  2002).  \emph{RHESSI} has nine coaxial germanium detectors, which record an X-ray emission from the full solar disk in a wide energy range (3 keV -17 MeV) with high temporal and energy resolutions as well as with a high signal sensitivity. Such characteristics allow a restoration of the 2D images and spectra in the X-ray band and provide very valuable data for investigation of the non-thermal emission of the solar flares. The X-ray emissions of the investigated flares were also recorded with the \emph{GOES} X-ray photometers. The Geostationary Operational Environmental Satellite (\emph{GOES}), operated by the United States National Environmental Satellite, Data, and Information Service, is fitted with two photometers, continuously recording full-disk integrated X-ray emissions in two energy bands 1-8 \AA\  and 0.5-4 \AA\ with 3 seconds temporal resolution (Donnelly et al. 1977). Both solar flares were also observed with the Extreme-ultraviolet Imaging Telescope (\emph{EIT}) installed on board the Solar and Heliospheric Observatory (\emph{SOHO}) (Delaboudiniere et al. 1995). The \emph{EIT} telescope provides full-disk images taken in four bands: 171 $\AA$, 195 $\AA$, 284 \AA\  and 304 \AA\  with 2.6 arcsec per pixel spatial resolution and the temperature range of the observed plasma is roughly $8 \times 10^4$ - $2 \times 10^6$ K.

\begin{figure*}[t!]
\includegraphics[angle=0,scale=0.26]{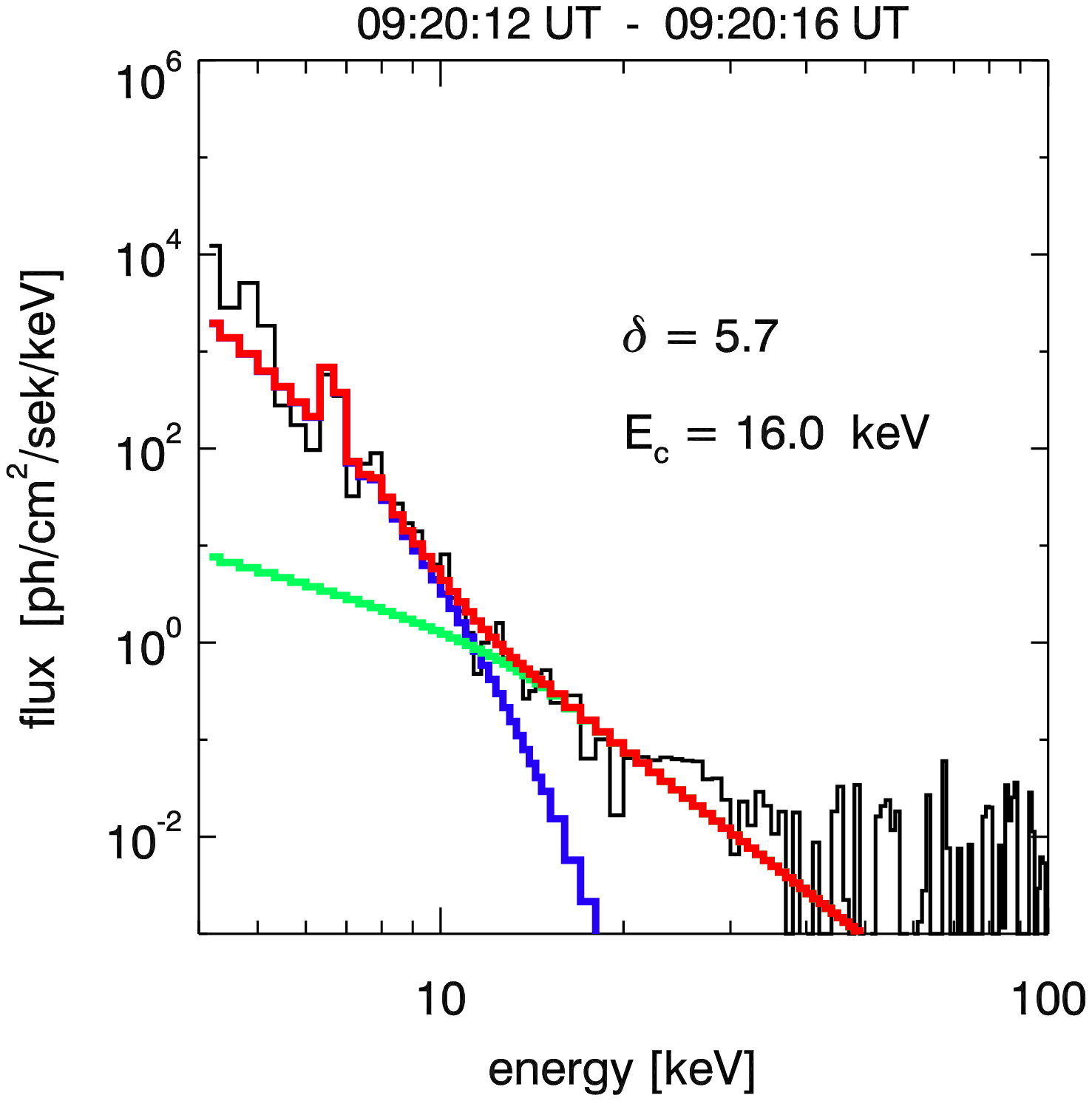}
\includegraphics[angle=0,scale=0.26]{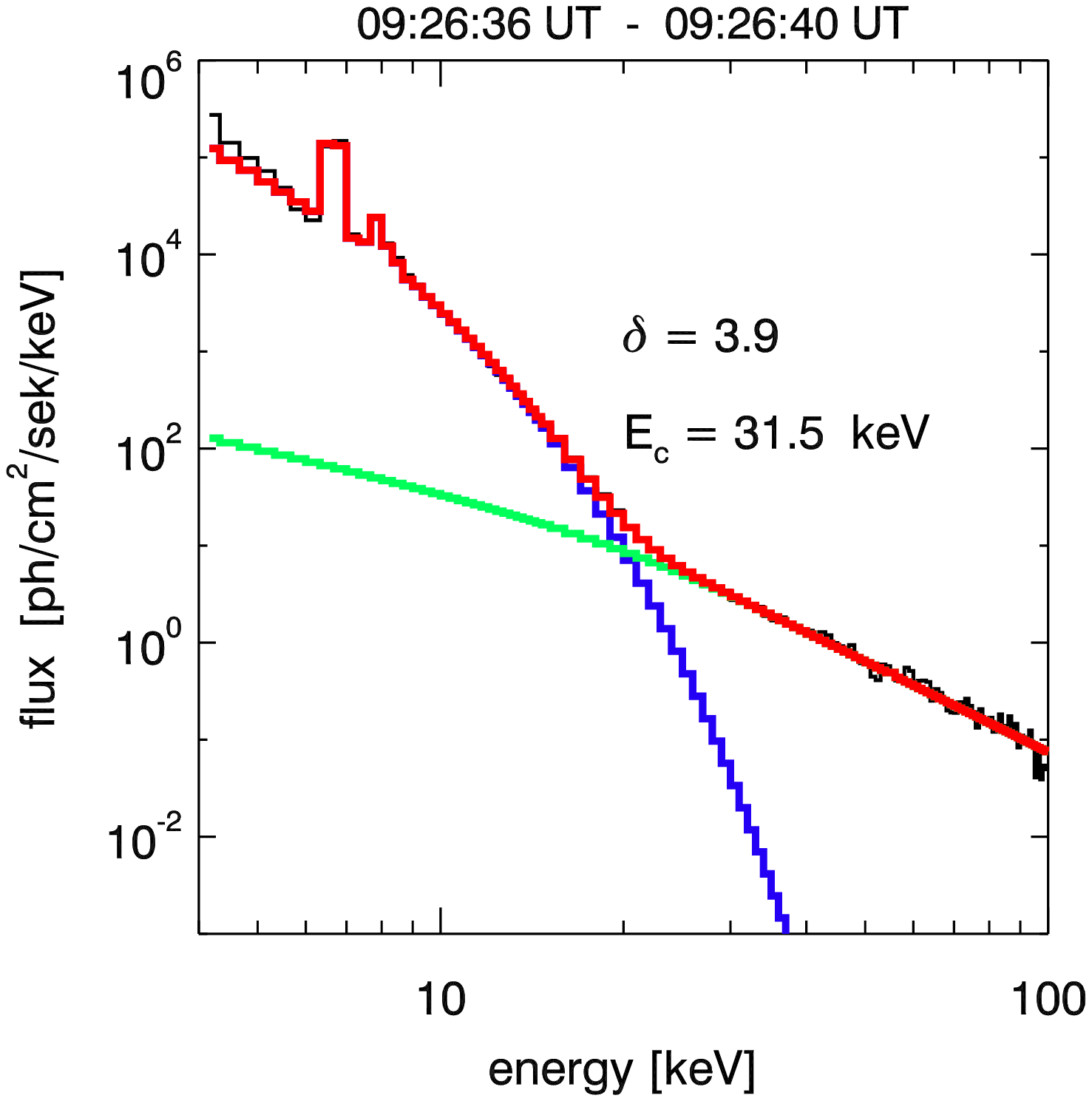}
\includegraphics[angle=0,scale=0.26]{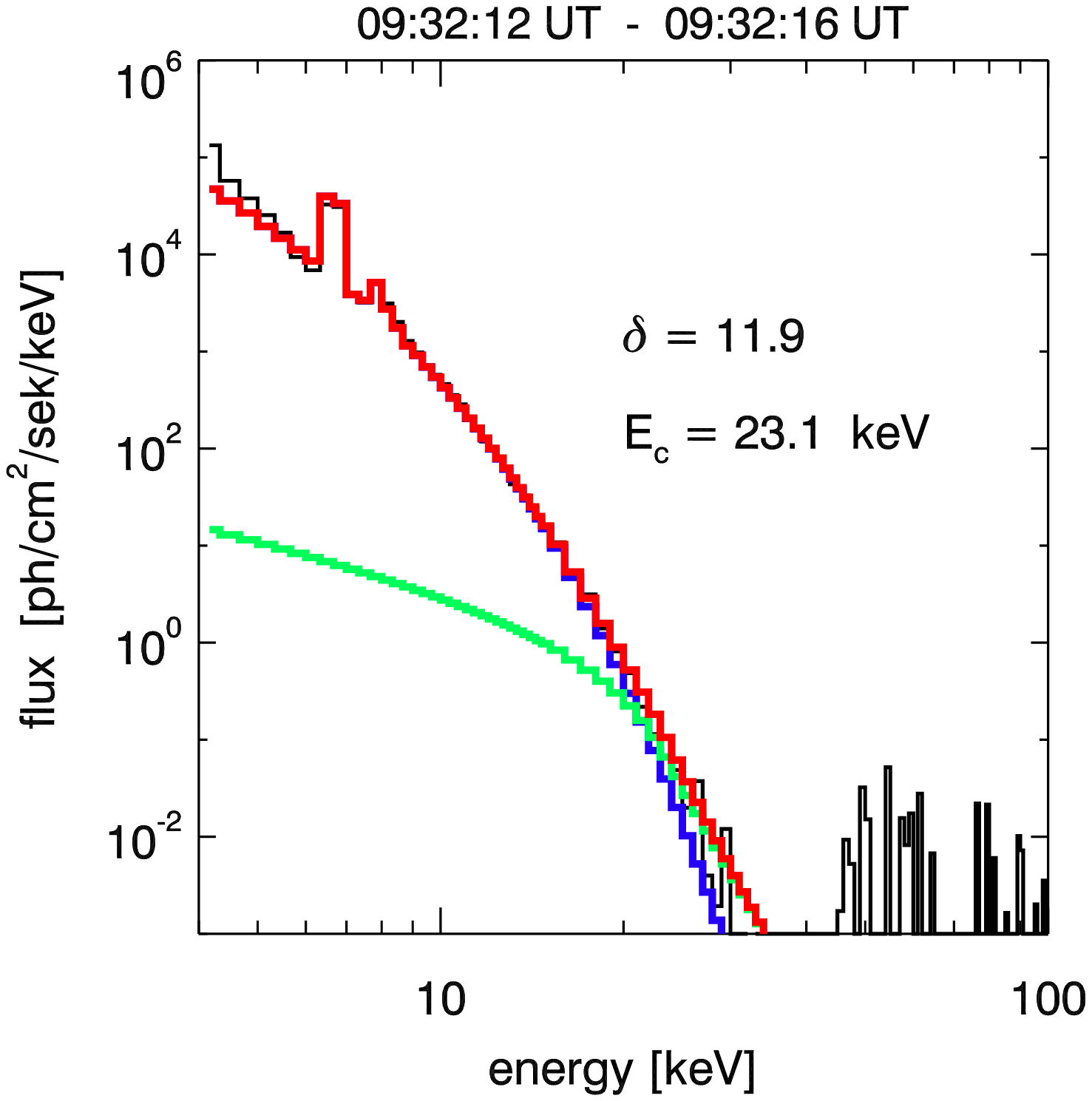}
\vspace{0.02 cm}
\caption{\emph{RHESSI} spectra taken before (left panel), during (middle panel) and after the impulsive phase (right panel) of the flare on September 20, 2002. The spectra were fitted with the single temperature thermal model (blue color) and thick-target model (green). The total fitted spectra are shown in red.
}
\end{figure*}

\subsection{September 20, 2002 solar flare}

\begin{figure}[t!]
\includegraphics[angle=0,scale=0.45]{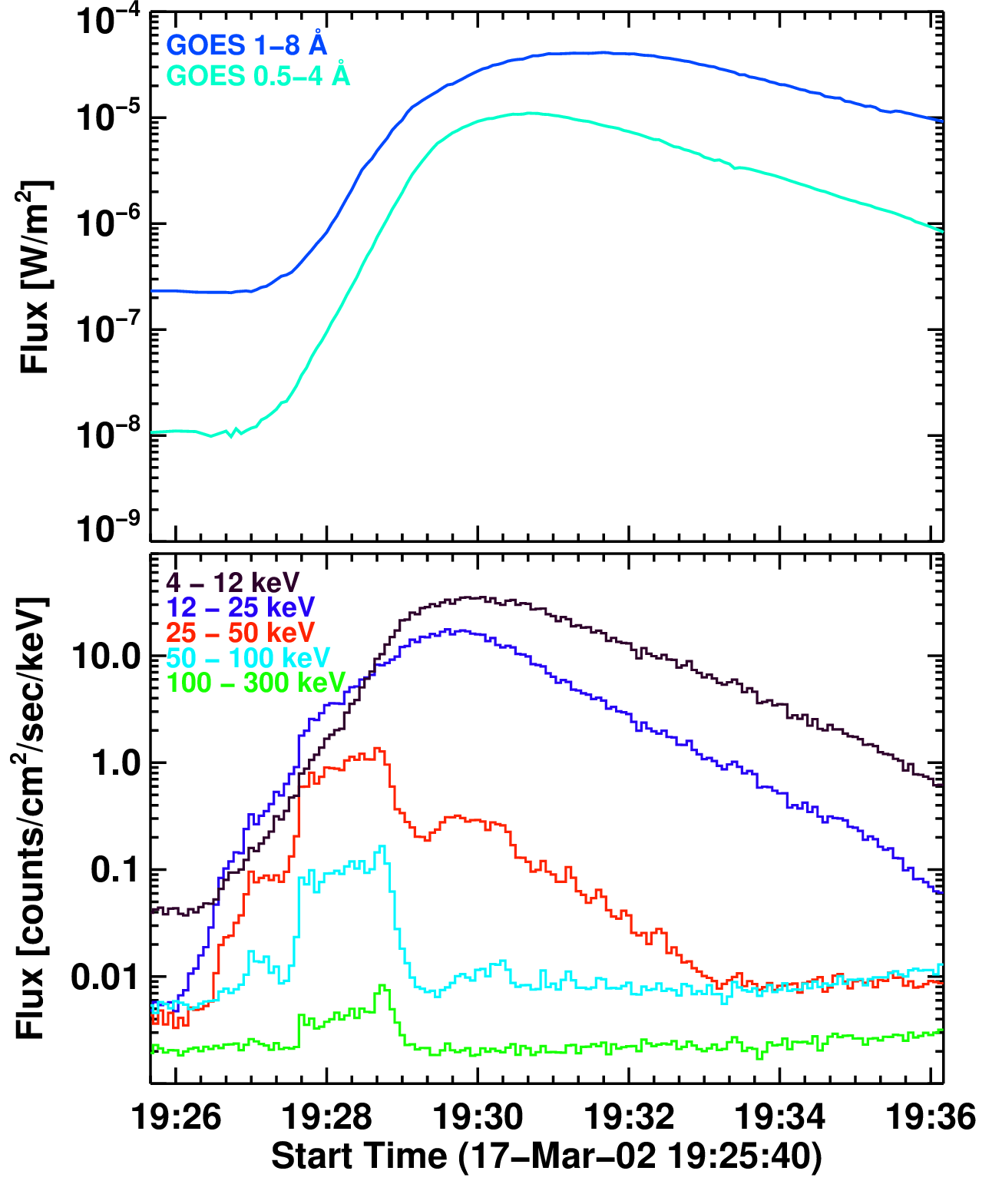}
\vspace{0.02 cm}
\caption{\emph{GOES} X-ray 0.5-4 \AA\  and 1-8 \AA\  light curves (upper panel) and \emph{RHESSI} light curves of five energy bands between 4 and 300 keV (lower panel) taken during the M.4 \emph{GOES} class solar flare on March 17, 2002.}
\end{figure}

\begin{figure*}[t!]
\includegraphics[angle=0,scale=0.45]{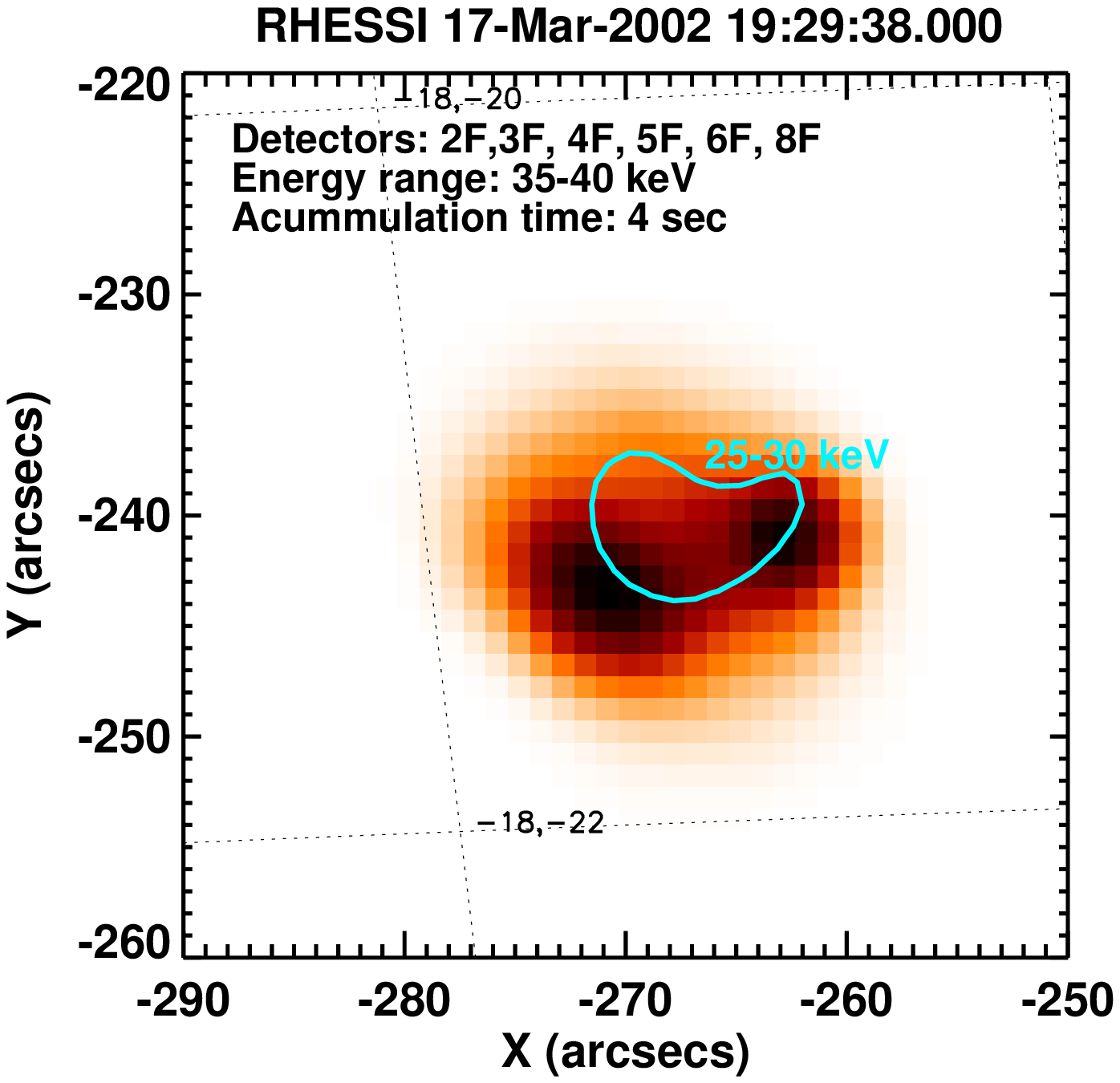}
\includegraphics[angle=0,scale=0.45]{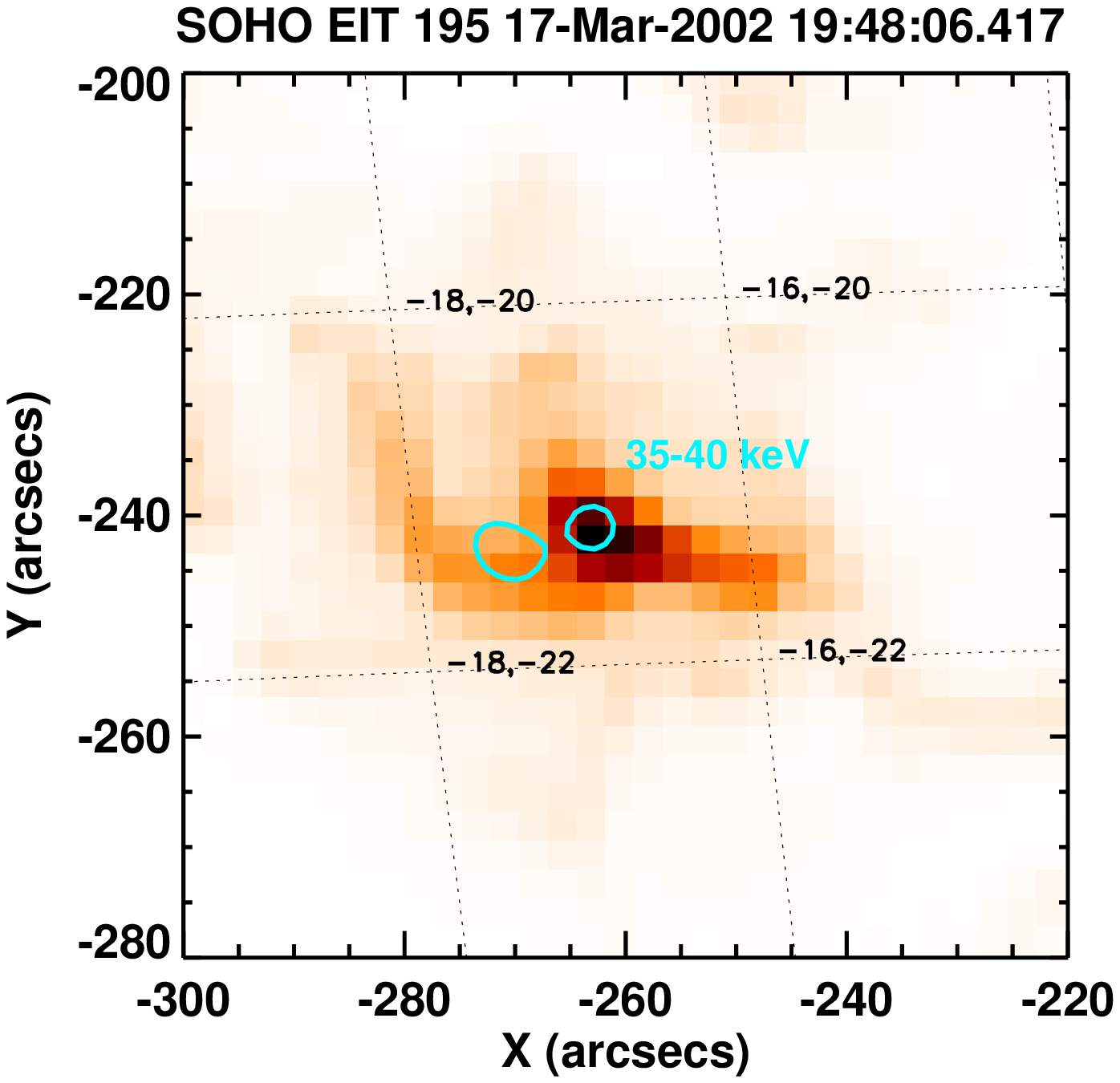}
\vspace{0.02 cm}
\caption{Images of the M4.0 \emph{GOES} class solar flare on March 17, 2002. Left panel: an image restored using the PIXON method in 35-40 keV energy band, signal was accumulated between 19:29:38 UT and 19:29:42 UT, at maximum of the impulsive phase (gray scale) over-plotted with \emph{RHESSI} $25-30$ keV image (contour). Right panel: \emph{SOHO/EIT} 195 \AA\ image taken at 19:48:06 UT, after the impulsive phase of the flare, over-plotted with \emph{RHESSI} 35 - 40 keV PIXON image (contour) registered at 19:29:38 UT.
}
\end{figure*}

The M1.8 \emph{GOES} class flare occurred in AR NOAA 10126 (S23E69) on September 20, 2002, its \emph{RHESSI} and \emph{GOES} light-curves are presented in Figure 1. The SXR (1-8 $\AA$) emission of the flare recorded by \emph{GOES} started at 09:18:15 UT, reached its maximum at 09:28:30 UT and was observed up to 10:00 UT. A harder emission recorded by \emph{GOES} (0.5-4 $\AA$) started to increase at the same time as the softer one but peaked one minute earlier at 09:27:30 UT. The impulsive phase of the flare recorded by \emph{RHESSI} in X-rays $\geq$ 25 keV started at 09:25:24 UT and had two maxima around 09:26 UT and 09:27 UT, respectively. In the 25-50 keV energy range, a small spike of emission was recorded between 09:24:16 UT and 09:24:32 UT. The SXR emission recorded by \emph{RHESSI} below 25 keV started to rise simultaneously with the SXR emission recorded by \emph{GOES} (see Paper I for more details).

Images of the flare were reconstructed using \emph{RHESSI} data collected with sub-collimators 2F, 3F, 4F, 5F, 6F, 8F and 9F, integrated over 8 second periods and PIXON imaging algorithm with 1 \emph{arcsec} pixel size (Metcalf et al. 1996, Hurford et al. 2002). The images revealed that SXR emission in 6-12 keV and intermediate 12-25 keV energy bands is coincident with the flare location. These observations also indicate that SXR emission recorded by \emph{GOES} during the early phase of the flare came from the analyzed event. The images registered in energies above 25 keV show two foot-points and loop-top source of a single flaring loop (see Figure 2). The images allow us to determine (using a method proposed by Aschwanden et al. 1999) the main geometrical parameters of the flaring loops, necessary for hydro-dynamic modeling. The cross-section of the loop $S=8.95\pm7.64\times10^{16}$ $cm^2$ was estimated as an area of the structure delimited by a flux level equal to 30\% of the maximum flux in the 25-35 keV energy range. The cross-sections of flaring loops of the both events analyzed in this paper were assumed to be constant. Half-length of the loop $L_0=9.31\pm1.13\times10^8$ $cm^2$ was estimated from a distance between the centers of gravity of the foot-points, assuming a semi-circular shape of the loop. Images obtained with the SOHO/EIT telescope in a 195 \AA\ band at 09:47:59 UT and 9:59:59 UT (after the impulsive phase of the flare) confirmed the single-loop structure of the flare (see Fig. 2, right panel).

\begin{figure*}[t]
\centering
\includegraphics[angle=0,scale=0.3]{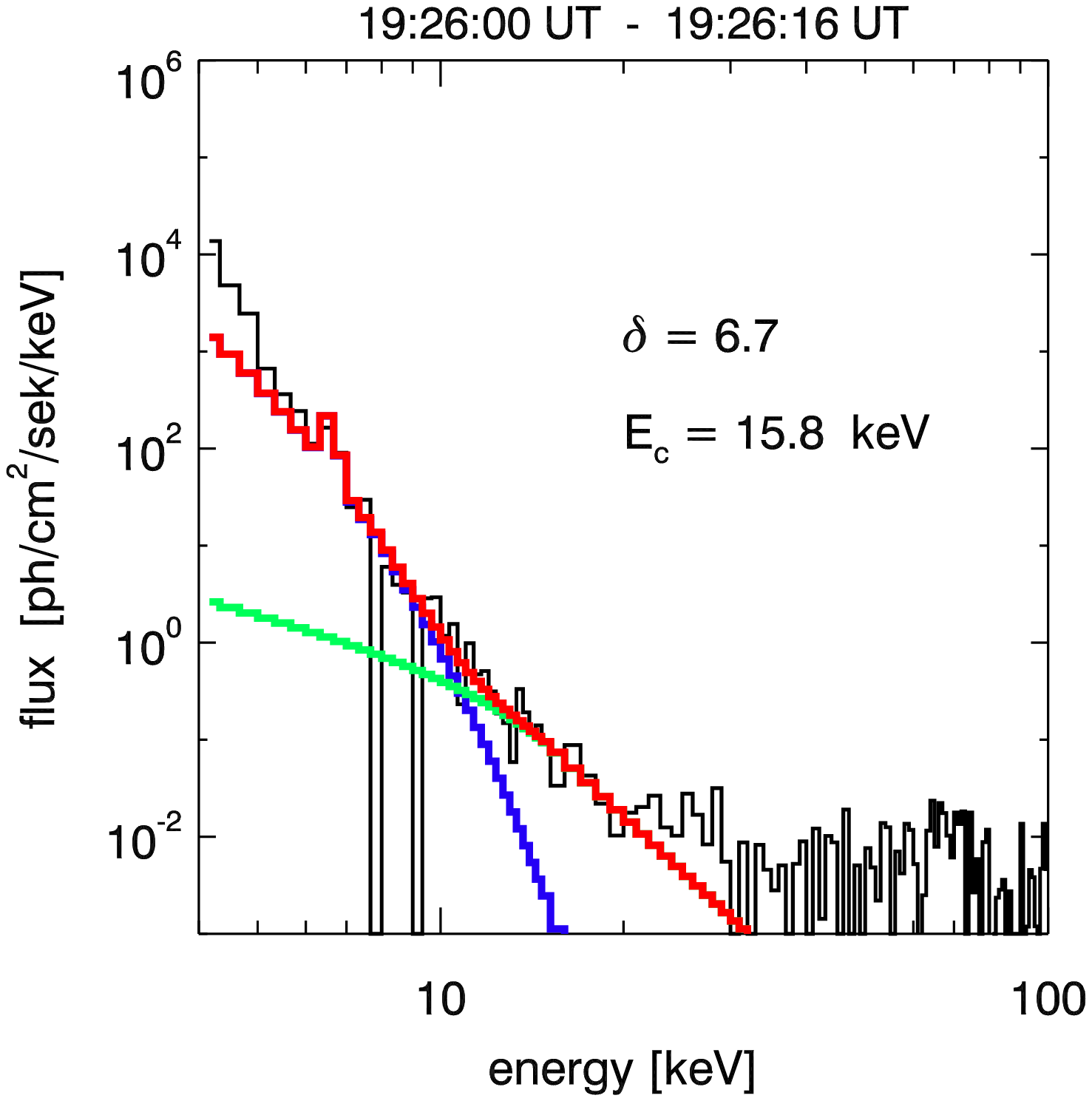}
\includegraphics[angle=0,scale=0.3]{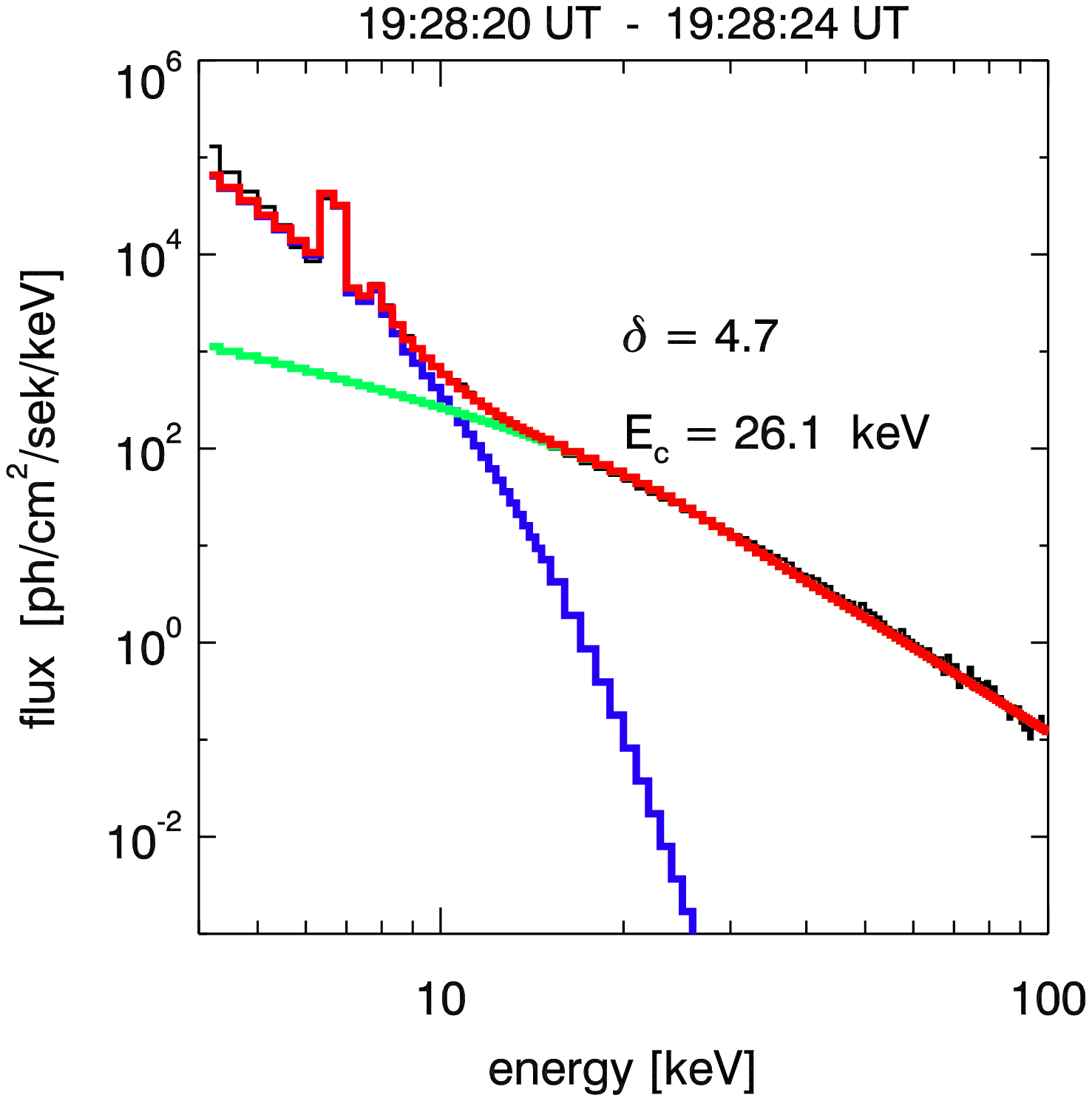}
\vspace{0.4 cm}
\caption{\emph{RHESSI} spectra taken before (left panel) and during (right panel) the impulsive phase of the flare on March 17, 2002. The spectra were fitted using a single temperature thermal model (blue color) and thick-target model (green). The total fitted model is shown in red.}
\end{figure*}

\subsection{March 17, 2002 solar flare}

The second investigated flare occurred as a M4.0 \emph{GOES} class event in the southeastern hemisphere in AR NOAA 9871 (S21E18) on March 17, 2002.  A magnetic class of the region was $\beta$, nevertheless it had already produced several C \emph{GOES} class solar flares. The SXR emission of the event started to increase slowly at 19:27 UT and showed a maximum at 19:31 UT, being observed up to 20:00 UT (\emph{GOES} X-ray light curves of the flare are shown in Figure 4). \emph{GOES} 1-8 $\AA$ flux has a background level of $6.29\times 10^{-6}$ $Wm^{-2}$ (C6.3). \emph{RHESSI} X-ray light curves of the flare taken in five energy bands are shown in Figure 4. The impulsive phase in X-rays above 25 keV started at 19:26:20 UT and had maximum at 19:29:40 UT. This event was also investigated in papers by Krucker \& Lin (2002) and Alexander \& Metcalf (2002).

The images of the flare were obtained using \emph{RHESSI} data collected with sub-collimators 2F, 3F, 4F, 5F, 6F and 8F integrated over 4 second periods and the PIXON imaging algorithm with 1 \emph{arcsec} pixel size. They showed a single, short and thick flaring loop (see Figure 5). The event was affected by pile-up effect during the analyzed time interval. The pile-up correction for spectra was applied, thus the introduced errors of the spectra parameters should be small compared with other uncertainties. In opposite, there is no pile-up correction for images and some low-energy (25-35 keV) emission of the loop was registered also in 35-40 keV band, increasing the relevant signal in between the feet of the loop. Thus, positions of the centers of gravity of the foot-points could be shifted closer towards each other causing some under-estimation of a semi-length of the loop.
The cross section of the loop $S = 4.21\pm1.66\times10^{17}$ $cm^2$ was estimated as an area of the structure delimited by a flux level equal to 30\% of the maximum flux in the 35-40 keV energy range, half-length of the loop was estimated as $L_0 = 4.42\pm 1.13\times 10^8$ $cm^2$ from a distance between the centers of gravity of the foot-points, assuming a semi-circular shape of the loop.

As in the case of the first event, the \emph{SOHO/EIT} telescope observed the active region before, during and after the flare. Unfortunately, only one saturated image registered during the flare was available. Two post-flare loops perpendicular to the flaring loop are recorded on the images taken in 195 \AA\ band after the flare (see right panel of the  Fig. 5).

\section{Modeling of the flares}

\begin{table*}[t!]
\caption{Main parameters of the analyzed flares applied in the calculations.} 
\label{table:1} 
\centering 
\begin{tabular}{c c c c c c c c} 
\hline\hline 
Event     & \multicolumn{2}{c}{Time of}& \emph{GOES}&Active & S              & $L_0$        & $P_0$       \\ %
date      &start     & maximum         & class      &region &                &              &             \\
          & [UT]     &[UT]             &            & AR    &[$10^{17} cm^2$]& [$10^{8}cm$] & [$dyn/cm^2$]\\
 \hline

20-Sep-02 &  9:21    &  9:28           & M1.8       &10126  &1.13            & 9.5          & 34.4        \\
17-Mar-02 & 19:24    & 19:31           & M4.0       &9871   &2.61            & 3.5          & 36.5        \\

\hline 
\multicolumn{8}{p{14cm}}{{\footnotesize $S$ and $L_0$ - cross-section and semi-length of the flaring loop, modified from the measured values (within error) in order to obtain the best conformity between modeled and observed
\emph{GOES} light curves; $P_0$ - pressure at base of transition region }} \\

\vspace{-0.6cm}
\end{tabular}
\end{table*}

Applied methods of data analysis and numerical modeling of the flares were similar to those presented in detail in Paper I, but we added here a set of procedures for automatic evaluation/optimizing of the low-energy cutoff of the electron distribution $E_c$ by comparison of the observed and calculated soft X-ray \emph{GOES} fluxes.

The HXR data were analyzed using the \emph{RHESSI} OSPEX package of the SolarSoftWare (SSW) package. The X-ray spectra of both flares were measured with 4 sec temporal resolution in 158 energy bands ranging from 4 to 300 keV and corrected for pulse pile-up, decimation, and albedo effects.
When the flux is low (at the beginning and end of the flare) the count rates in some energy bins can be negative as a result of the background subtraction due to low signal to noise ratios. In order to keep the count rates
positive (at least at 4 - 20 keV energy range) we increased the accumulation times. As already presented in Paper I the spectra were fitted using single temperature thermal plus thick-target models (vth + thick). The thermal model was defined by single temperature and emission measure of the optically thin thermal plasma, the thick-target model was defined by the total integrated NTE flux $F$, the power-law index of the electron energy distribution $\delta$, and the low-energy cutoff of the electron distribution $E_c$. The \emph{RHESSI} spectra were fitted using a forward and backward automatic fitting procedure at  all times, starting from a moment when the non-thermal component was strong and clearly visible. The obtained values of the fitted parameters were additionally controlled and corrected, if necessary. This procedure causes that usually all fitted parameters evolve in a quasi continuous manner. The averaged non-flare background spectra were removed before the fitting procedure. The background spectra for energies below 50 keV were accumulated and averaged from pre-flare periods between 09:00 and 09:06 UT for September 20, 2002 and between 19:00 and 19:12 UT for March 17, 2002 flares, respectively. In the case of the September 20, 2002 flare we used a linear interpolation between the time intervals before and after the impulsive phase for energies above 50 keV.

\begin{figure*}[t!]
\includegraphics[angle=90,scale=0.5]{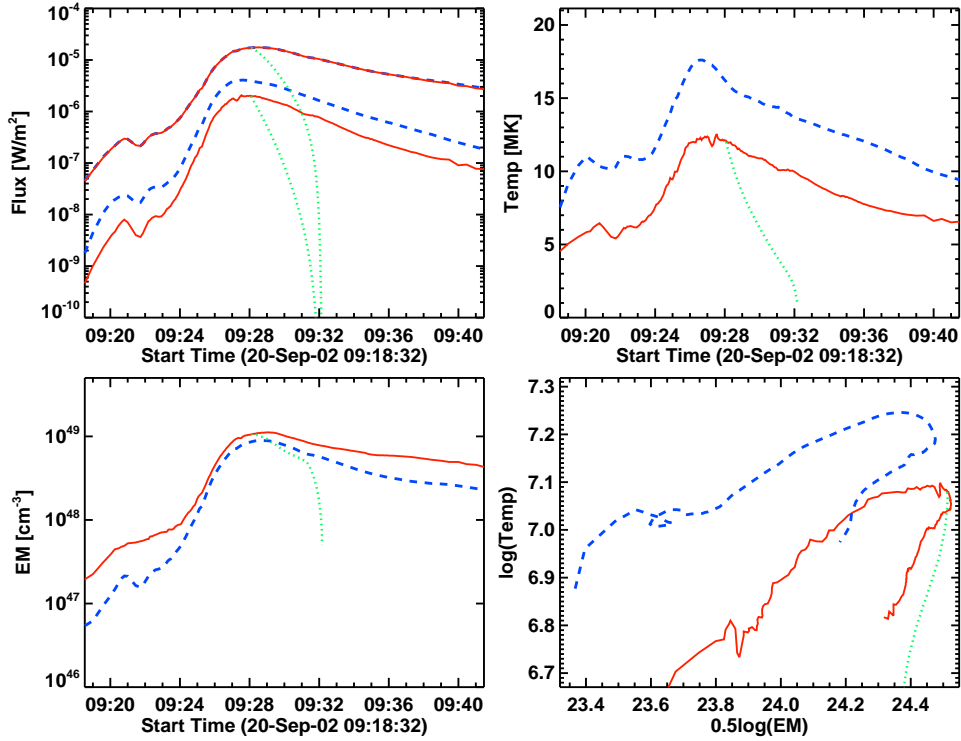}
\vspace{0.02 cm}
\caption{Results of the modeling of the September 20, 2002 solar flare. Upper left panel: blue dashed lines - observed \emph{GOES} fluxes in 0.5-4 \AA\  (lower curve) and 1-8 \AA\  (upper curve) energy bands; red lines - calculated \emph{GOES} fluxes; green (dotted) curves - calculated \emph{GOES} fluxes without any heating after 09:28:00 UT. Upper right panel and two lower panels: temperature, emission measure and diagnostic diagram log(T) vs. 0.5log(EM), respectively. Blue lines - values calculated using \emph{GOES} data; red lines - values modeled; green lines - values modeled without any heating after 09:28:00 UT (switched off).
}
\end{figure*}

\begin{figure}[t!]
\includegraphics[angle=0,scale=0.50]{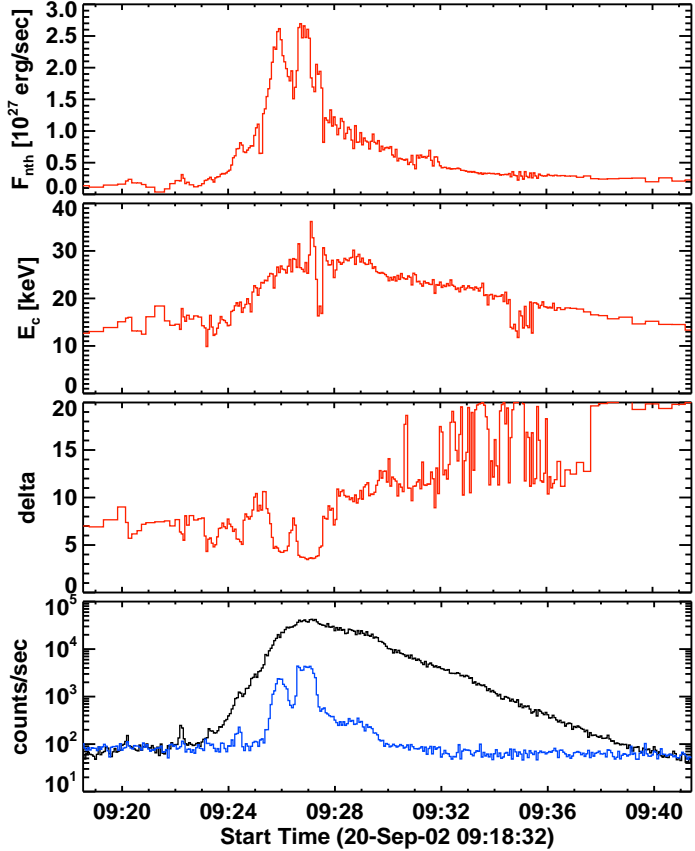}
\vspace{0.02 cm}
\caption{Time evolution of the thick-target model parameters calculated for the September 20, 2002 solar flare using \emph{RHESSI} registered fluxes. From top to bottom: energy flux of non-thermal electrons $F_{nth}$, cutoff energy $E_c$, $\delta$ index of the energy spectrum and HXR fluxes: 12-25 keV (black line) and 25-50 keV (blue line).
}
\end{figure}

The fundamental assumption of our work was that only non-thermal electron beams derived from \emph{RHESSI} spectra delivered energy to the flaring loop (via the Coulomb collisions with the plasma filling the loop). The deposition of the energy by NTEs was modeled by us using an approximation given by Fisher (1989). The hydro-dynamic evolution of the flaring plasma was modeled with the modified Naval Research Laboratory Solar Flux Tube Model code (Mariska et al. 1982, 1989, see Paper I and Falewicz et al. (2009) for details). Initial, quasi-stationary pre-flare models of the flaring loops were built using geometrical (semi-length $L_0$, cross-section $S$) and thermodynamic (initial  pressure  at  base  of  transition  region $P_0$, temperature, emission measure, mean electron density and GOES-class) parameters estimated from \emph{RHESSI} and \emph{GOES} data. The geometric parameters of the loop $L_0$ and $S$ were evaluated under the assumption that an observational error of the position of the observed structure is of the order of one pixel. In a course of the calculations both semi-lengths and cross-sections of the loops were refined (in a range of the error only) in order to obtain the best conformity between theoretical and observed \emph{GOES} light curves (Table 1 presents values of $S$, $P_0$ and $L_0$ used in calculations).

For each time step of numerical modeling we estimated the momentary heating rate of the plasma along the loop (i.e. an amount and a distribution of the deposited energy) using Fisher's heating function, employing suitably thick-target parameters $F$, $\delta$ and $E_c$ of the NTEs beams derived from fitted consecutive \emph{RHESSI} spectra. Next we evaluated a momentary distribution of the hydro-dynamic parameters of the plasma and the resulting calculated \emph{GOES} fluxes.

An estimation of the total energy carried by the NTEs is very sensitive to the evaluated low energy cutoff of the electron spectrum $E_c$, because of the power-law nature of the electron energy distribution. In other words a value of the low energy cutoff $E_c$ determines an amount of energy delivered to the loop. A variation in the $E_c$ value of just a few keV can add/remove a substantial amount of energy to/from the modeled system/flare, so $E_c$ must be selected with the greatest care. However, we found that acceptable conformities of the calculated and observed fluxes could be obtained using various values of $E_c$ in the range from $\sim$5 to $\sim$30 keV for all times in both analyzed events. This non-uniqueness could be limited using an independent energetic condition, like an observed 1-8 \AA\ \emph{GOES} flux. Indeed, for each time step we carefully adjusted the $E_c$ value in order to achieve conformity of the observed and modeled \emph{GOES} fluxes in 1-8 \AA\ band.

\begin{figure*}[t!]
\includegraphics[angle=90,scale=0.5]{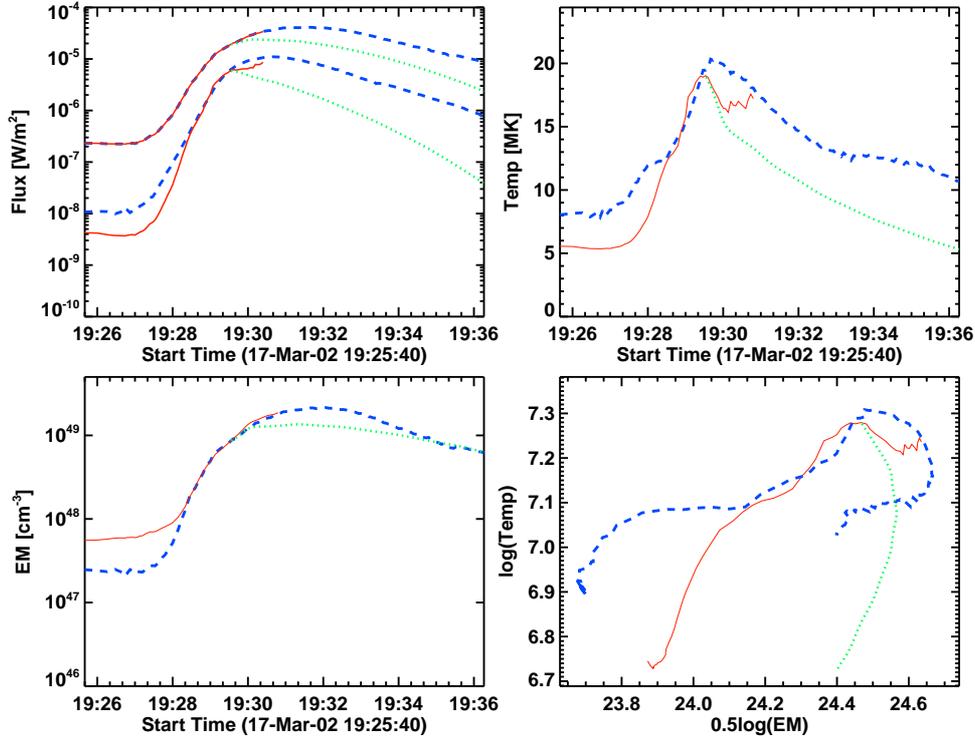}
\vspace{0.02 cm}
\caption{Results of the modeling of the March 17, 2002 solar flare. Upper left panel: blue dashed lines - observed \emph{GOES} fluxes in 0.5-4 \AA\  (lower curve) and 1-8 \AA\  (upper curve) energy bands; red lines - calculated \emph{GOES} fluxes; green (dotted) curves - calculated \emph{GOES} fluxes without any heating after 19:29:10 UT. Upper right panel and two lower panels: temperature, emission measure and diagnostic diagram log(T) vs. 0.5log(EM), respectively. Blue lines - values calculated using \emph{GOES} data; red lines - values modeled; green lines - values modeled without any heating after 19:29:10 UT (switched off).}
\end{figure*}

As two illustrative examples of the relation between $E_c$ and heating rate of the loop we could show a pre-heating phase of March 17, 2002 solar flare and decay phase of the September 20, 2002 solar flare. For the March 17, 2002 solar flare at 19:26:00-19:26:16 UT (see Figure 6, left panel), assuming three various values of $E_c$: 13.7 keV, 14.7 keV and 15.7 keV we obtained the relevant \emph{GOES} classes of the emission: B2.45, B2.33 and B2.26 and NTEs energy fluxes: $1.66\times10^{26}$ $erg/sec$, $1.19\times10^{26}$ $erg/sec$ and $8.73\times10^{25}$ $erg/sec$, respectively. The final value of the $E_c$, giving a conformity of the calculated and observed \emph{GOES} fluxes in 1-8 \AA\ band (B2.25 \emph{GOES} class) was equal to 15.8 keV and the NTEs energy flux was equal to $8.56\times10^{25}$ $erg/sec$. During the decay phase of the September 20, 2002 solar flare at 09:32:12-09:32:16 UT (see Figure 3, right panel), for three values of $E_c$: 18.2 keV, 20.2 keV and 22.2 keV we obtained the relevant \emph{GOES} classes of the emission: M1.14, M1.04 and C9.96 and NTEs energy fluxes: $4.38\times10^{27}$ $erg/sec$, $1.55\times10^{27}$ $erg/sec$ and $6.09\times10^{26}$ $erg/sec$, respectively. The final value of the $E_c$ was equal to 23.1 keV and the NTEs energy flux was equal to $4.06\times10^{26}$ $erg/sec$ for the observed \emph{GOES} class C9.86. The problem of the influence of the $E_c$ variations onto the resulting light-curves and classes of the flares was thoroughly investigated already by Falewicz et al., (2009). Figures 3 and 6 show examples of the \emph{RHESSI} fitted spectra where low-energy cutoff $E_c$ were adjusted in order to equalize synthesized and observed \emph{GOES} fluxes in 1-8 \AA\ band. The total energy fluxes calculated for the whole event on September 20, 2002 and for the analyzed part of the March 17, 2002 flare are in conformity with the fluxes evaluated by various authors for solar flares having comparable GOES-classes and geometrical parameters (e.g. McDonald et al., 1999; Saint-Hiliare \& Benz, 2005).

\begin{figure}[t!]
\includegraphics[angle=0,scale=0.50]{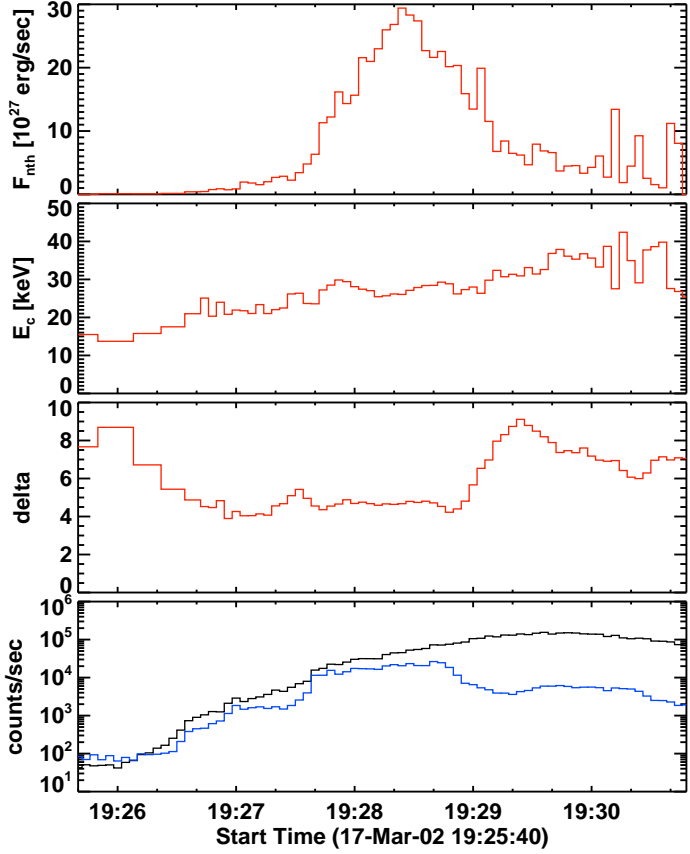}
\vspace{0.02 cm}
\caption{Time evolution of the thick-target model parameters calculated for the March 17, 2002 solar flare using \emph{RHESSI} recorded fluxes. From top to bottom: energy flux $F_{nth}$, cutoff energy $E_c$, $\delta$ index of the energy spectrum and HXR fluxes: 12-25 keV (black line) and 25-50 keV (blue line).
}
\end{figure}

The synthesized \emph{GOES} 1-8 \AA\ light curves of the flaring loops follow closely the observed ones for both events, which directly results from our assumptions. Unfortunately, the correspondences between the observed and calculated fluxes in the 0.5-4 \AA\ band are not so ideal (see Figures 7 and 9). The inconsistency could be attributed to: relative simplicity of the applied numerical model, errors in \emph{RHESSI} spectra restoration, crude estimation of the initial loops' conditions and possible problems of a \emph{GOES}'s calibration. However, because the calculated 0.5-4 \AA\ light curves did not differ too much from the observed ones, it seems that our model simulates the main physical processes in the right way.

The early (pre-heating) phase of the flare observed on September 20, 2002 was already modeled by us in Paper I. Presently we modeled the whole evolution of the same flare assuming that necessary small heating on the decay phase is also caused by NTEs (we discuss this point in detail in the Discussion and Conclusion section). We started modeling of the flare a few minutes before the impulsive phase, proceeding through the maximum of the flare far into the gradual phase of the flare. Figure 8 presents time variations of the electron beam (thick target) parameters and \emph{RHESSI} fluxes of the flare. NTEs beams characterized by these parameters, provided the loop with an amount of energy fully sufficient to power the fluency of the \emph{GOES} emission observed before and during the impulsive phase, and also during the decay phase of this flare. Low-energy cutoff $E_c$ varied during the flare between 9.9 keV and 36 keV, electron spectral index $\delta$ varied between 3.5 and $\sim$20.0, while energy flux of NTEs $F_{nth}$ ranged from $3.7\times 10^{25}$ $erg/s$ to $2.7\times 10^{27}$ $erg/s$. The small differences between models presented in previous and present papers noticeable during the early phase of the flare are caused by small refinements of the initial geometrical and physical parameters of the flare.

We also modeled the pre-heating and impulsive phases of the solar flare observed on March 17, 2002. However, we were not able to reproduce the decay phase of this flare, apparently due to a complicated magnetic structure of the event. During the decay phase when a double-loop structure was well visible, our single-loop numerical model was not relevant.  While the volume of the heated plasma inside two flaring loops undoubtedly increased, the energy delivered by the non-thermal electrons to the flare increased accordingly in order to cover total losses of both loops. As a result, in our single-loop model we obtained an increased chromospheric evaporation (due to the increased energy input) which caused an excessive emission of SXR.  In order to equalize observed and modeled SXR fluxes, $E_c$ should be (in our model) of the order of about 300 keV - obviously a non-physical value. Figure 10 presents time variations of the parameters of the electron beam (in thick target approximation) and \emph{RHESSI} fluxes of this flare. Low-energy cutoff $E_c$ changed during the rise phase and maximum of the flare between 13.7 keV and 42 keV, electron spectral index varied between 3.9 and 9.0, while energy flux of non-thermal electrons ranges from $1.9\times 10^{25}$ $erg/s$  to $2.9\times 10^{ 28}$ $erg/s$.

Most of the solar flares show during the impulsive phase a typical pattern of the variations of the observed spectral index ($\delta$): soft-hard-soft (see, e.g. Grigis \& Benz 2004). In the case of the solar flare observed on September 20, 2002 this pattern is well visible between 19:26:00 UT and 19:29:30 UT, while for the March 17, 2002 event it is noticeable between 19:25:00 UT and 19:28:00 UT (see Figures 8 and 10, respectively). Obtained variations of the $E_c$ could be caused by temporal variations of the processes in primary regions of the magnetic energy release and acceleration of the non-thermal electrons. Additionally, all peaks registered  in the 12-50 keV energy range are related to local increases in energy flux of the non-thermal electrons and to local increases in heating of the loop.

We can conclude that for a single-loop flare on September 20, 2002 we were able to restore the observed temporal variations of the SXR fluxes emitted during the pre-heating, maximum and decay phases using only energy carried by non-thermal electron beams derived from the observed HXR spectra. Under the same assumption we also restored the observed temporal variations of the SXR fluxes emitted during the pre-heating and maximum phases of the solar flare on March 17, 2002 when it had a single-loop structure.

\section{Discussion and Conclusions}

In our work we showed that an SXR emission observed for single-loop flares from the pre-heating up to decay phases could be fully explained by electron beam-driven evaporation only and without any ad hoc assumptions concerning any other auxiliary heating mechanisms, while all energy necessary to explain the observed SXR emission and dynamics of the flaring plasma could be derived from observed HXR spectra. Our result extends the standard model of the SXR and HXR relationship to the very early and decay phases of solar flares. The result also indicates that the process of electron's acceleration can occur during both: early and decay stages of the flares, well before and well after the impulsive phase.

For a single-loop flare on September 20, 2002 we were able to restore the observed temporal variations of the SXR fluxes emitted during the pre-heating, maximum and decay phases using only energy carried by non-thermal electron beams derived from the observed HXR spectra. In the case of the solar flare observed on March 17, 2002 we were not able to achieve a satisfactory compliance between temporal variations of the modeled and observed SXR fluxes for the late (gradual) phase of the flare recorded after 19:31:53 UT. However, images of the flare obtained during that phase of the event (see Figure 5, right panel) show two (or more) bright and probably interacting loops. While our 1D numerical model of the flaring loop is not relevant to a multi-loop flare, we limited our calculations of the flare evolution to the initial phase of the flare only, ceasing the calculations at the moment corresponding to the real solar flare evolution at 19:31:30 UT, just before the maximum of the SXR emission (see Figure 9).

\subsection{Temporal variations of $E_c$}

The modeled temporal variations of the cutoff energy $E_c$ of the flares observed on September 20, 2002 and March 17, 2002 (see Figures 8 and 10, respectively) agree well with estimations of the $E_c$ ranges made already by various authors (see e.g. Holman et al. (2003), Sui et al. (2007) and Han et al. (2009)).  Additionally, our method of adjustment of the low-energy cutoff $E_c$ in order to equalize synthesized and observed \emph{GOES} fluxes in 1-8 \AA\ band can be considered as a new method of $E_c$ determination. In the case of the September 20, 2002 solar flare cutoff energy during the modeled pre-heating phase of the flare varied between 12 keV and 18 keV, gradually increasing up to about 30 keV during the maximum of the flare, decreasing back to about 15 keV during the gradual phase of the flare.  In a case of the solar flare observed on March 17, 2002 the cutoff energy $E_c$ started from about 15 keV during the pre-heating phase and next gradually increased up to about 40 keV during the final stages of the flare. The temporal variations of the $E_c$ evaluated by us are similar to variations of this parameter already reported by several authors (for example: Warmuth et al., 2009a; Wermuth et al., 2009b; Sui et al, 2007; Holman et al., 2003).

The variations of the $E_c$ could reflect temporal variations of the processes in the primary energy source and/or acceleration region but it could also be an effect of the modeling only. What is more, the accuracy of the obtained results was limited by errors of the estimation of the initial loop physical and geometrical parameters, errors in \emph{RHESSI} spectra restoration and GOES calibration, the simplicity of 1D hydro-dynamical modeling, and applied single loop approximation.
The non-thermal fits in the decay phase had quite small formal errors of fitting parameters, being of the order of 1\% ($\delta$ and $E_c$). The statistical errors of the same parameters, estimated using their values calculated in the numerical model, were of the order of $15-20\%$ for $\delta$, and $10-15\%$ for $E_c$.
Taking into account the relative simplicity of the numerical code for applied the flaring loops we achieved a startling overall concordance between temporal variations of the modeled and observed SXR fluxes for the flares. The concordance is especially good for the solar flare observed on September 20, 2002, undoubtedly due to its simple, single-loop structure.

\subsection{Fitting of the \emph{RHESSI} spectra}

Our calculations made using the single temperature thermal plus thick-target models show that the modest non-thermal part of the energy spectrum reveals energy carried by NTEs sufficient to power observed temporal changes of the soft X-rays emitted by the flare and to balance conductive as well as radiative energy losses of the flaring loops. However, the spectra registered by \emph{RHESSI} X-ray photometers, particularly spectra recorded during pre-heating and gradual phases of the solar flares, could also be reasonably fitted with the thermal model only, leading to an acceptably low value of the $\chi^2$ estimator. Thus for comparative purposes we also fitted the spectra recorded during the pre-heating phases with the purely thermal model. The temperatures obtained for both events seem to be quite high with respect to the temperatures evaluated from the \emph{GOES} data and make us more confident in the interpretation with the presence of the non-thermal electrons at least well before the impulsive phases of the solar flares. For September 20, 2002 we obtained a temperature of 20.3 MK, much higher than the temperature of the $\sim$8 MK evaluated using \emph{GOES} data, which seems to be too high to be real at the very early stage of the flare evolution. It is worth stressing that the differences between \emph{RHESSI} and \emph{GOES} temperatures mentioned in the literature (i.e. McTiernan, 2009; Raftery et al., 2009) are of the order of 2-6 MK, when the \emph{GOES} emission measure is typically 50-100 times greater than the \emph{RHESSI} emission measure. For the same time the difference of the \emph{RHESSI} and \emph{GOES} emission measures was of the order of 10 only, thus the expected difference of the temperatures should be even less than 6 MK. In our model the difference of the conductivity flux and heating flux by non-thermal electrons is of the order of 100-10000. Even assuming a much higher plasma temperature of 20 MK, the non-thermal heating remains still more than 20 times greater than the conductive flux.

During the decay phase of the analyzed flare the differences between temperatures evaluated using the \emph{RHESSI} and \emph{GOES} data were much lower, and there is no clear indication of a presence of a non-thermal component in the spectra. Although the pure thermal model can formally fit the observed spectra it cannot explain the observed fluency of the \emph{GOES} light-curves. Turning off any heating leads to a dramatic decrease of the calculated \emph{GOES} fluxes. So, at this stage of the flare a continuous delivery of some energy to the plasma is obvious and necessary.

The thermal evolutions of the flaring plasma of both events are presented in so-called diagnostic diagrams (Jakimiec et al., 1992).
On the diagnostic diagram, where the emission measure is given on the horizontal axis while plasma temperature
is given of the vertical axis, the efficiency of the plasma heating during the
decay stage of the flare is reflected by a slope of its evolutionary curve.
If the heating of the flaring loop is abruptly switched off, the slope of the evolutionary curve is roughly 2. In opposite, if the energy losses are still fully covered by energy input, the loop evolves quasi-stationary (so called quasi-stationary-state or QSS) and the slope of the evolutionary curve is equal to about 0.5. Figure 7 (right lower panel) shows the diagnostic diagram for the September 20, 2002 solar flare. The observed and modeled evolutionary curves have a slope less than 2, which indicates that some heating was present during the decay phase of the event. A green dotted curve shows an evolutionary track calculated in the special case of the abruptly switched off heating after 09:28:00 UT. It has a slope very close to 2. The shift of the observed and calculated evolutionary tracks is caused mainly by under-estimation of the modeled temperatures of the plasma.
For the March 17, 2002 solar flare (see Figure 9, lower right panel) the slope of the observed evolutionary track is much lower than 2, which indicates that some heating was present also during the decay phase of that event. Once again, the green dotted curve shows an evolutionary track calculated in the special case of the abruptly switched off heating after 19:29:10 UT, just after its impulsive phase.

\subsection{The non-thermal electrons during the decay phase}

We assumed here that during the gradual phase of the investigated flare energy is delivered by non-thermal electrons and thus we were able to evaluate the necessary amount of this energy. The evaluated energy is small (compared to e.g thermal energy contained in the flaring loop) but it is fully sufficient to fulfil the energy budget of the plasma during the decay phase of the flare. While this energy is small, at least two causes can make the non-thermal component of the spectrum less clear or even invisible at first glance of a decay phase. The first cause is masking /or shading/ of the weak non-thermal emission by a strong thermal one while during the decay phase the hot and dense plasma emits a large amount of SXR, much stronger than HXR in the same energy band. Secondly, no apparent symptoms of the  presence of the non-thermal particles does not necessarily mean that the electrons do not exist, while very high  \emph{RHESSI} sensitivity can be still too low to reveal clearly their emission. Indeed,  shown by Brosius \& Holman (2009), the non-thermal HXR emission associated with the electron beam powerful enough to heat chromospheric plasma to the temperature and emission measure observed by \emph{RHESSI} should be well below \emph{RHESSI}'s detection threshold. This is the case presented in Figure 3 (right panel).

\subsection{Final Conclusions}

Under the assumption that non-thermal electrons are the only source (i.e. carrier) of the energy which heated plasma during the whole flare, we calculated the energy flux contained in a non-thermal electrons beam and we showed that it was fully sufficient to fulfil energy budget of the plasma during the decay phase of the September 20, 2002 flare. What is more, a purely thermal model cannot explain the observed fluency of \emph{GOES} light-curves.

Thus, we show that in both analyzed flares the whole energy necessary for heating the flaring loops during the pre-impulsive and impulsive phases as well as during the post-impulsive (gradual) phase of one flare could be delivered by non-thermal electron beams only, whose parameters are restorable using \emph{RHESSI} and \emph{GOES} observational data. There is no need to model the pre- and post-impulsive phases of those flares using any additional ad-hoc heating mechanisms other than heating by non-thermal electrons.

\section{Acknowledgments}
The authors acknowledge the \emph{RHESSI} consortium and are grateful to the anonymous referee for constructive comments and suggestions, which have proved to be very helpful in improving the manuscript. MS was supported by the European Community's Seventh Framework Programme (FP7/2007-2013) under grant agreement no. 218816 (SOTERIA project), and by the Polish Ministry of Education and Science Grant No. N203 381736.

\clearpage

\end{document}